\documentclass[aps,prd,10pt,superscriptaddress,preprint,nofootinbib]{revtex4-1}
\usepackage{graphicx, epsfig} 
\usepackage{amsmath,amssymb,amsfonts,dsfont,mathrsfs,amsthm,mathtools}
\usepackage{bm} 
\usepackage{color}
\usepackage[usenames]{xcolor}
\usepackage{hyperref}
\usepackage{siunitx}
\hypersetup{colorlinks=true,urlcolor=blue,linkcolor=magenta,citecolor=blue,filecolor=blue}
\usepackage[normalem]{ulem}
\usepackage{array}
\usepackage{hyperref}
\usepackage{booktabs}
\usepackage{blindtext}

\usepackage{bigints}

\newcommand{\diff}[1]{\text{d}#1}

\newcommand{\Lag}{\mathscr{L}}

\newcommand{\ri}{\textrm{i}}
\newcommand{\rd}{\textrm{d}}

\begin{document}

\title{Planar AdS multi-NUT spacetimes and Kaluza-Klein multi-monopoles}

\author{Crist\'obal \surname{Corral}}
\email{cristobal.corral@uai.cl}
\affiliation{Departamento de Ciencias, Facultad de Artes Liberales, Universidad Adolfo Ib\'{a}\~{n}ez, Avenida Padre Hurtado 750, 2562340, Vi\~{n}a del Mar, Chile}

\author{Cristi\'an \surname{Erices}}
\email{cristian.erices@ucentral.cl}
\affiliation{Centro de Investigación en Ciencias del Espacio y Física Teórica
(CICEF), Universidad Central de Chile, La Serena 1710164, Chile}

\author{Daniel \surname{Flores-Alfonso}}
\email{dafa@azc.uam.mx}
\affiliation{Departamento de Ciencias B\'asicas, Universidad Aut\'onoma Metropolitana -- Azcapotzalco, Avenida San Pablo 420, Colonia Nueva El Rosario, Azcapotzalco 02128, Ciudad de M\'exico, Mexico}

\author{Benjam\'in \surname{Hernandez}}
\email{bhernandez2019@udec.cl}
\affiliation{Departamento de F\'isica, Universidad de Concepci\'on, Casilla, 160-C, Concepci\'on, Chile}

\begin{abstract}
In higher-dimensional Einstein-AdS gravity, it is well known that planar and static anti-de Sitter black holes can be endowed with multiple rotation parameters via a large-gauge transformation. However, a similar prescription fails when multiple NUT parameters are added, thereby obstructing the study of holographic properties with more than one NUT charge. To pave the way towards this direction, we construct explicit planar AdS spacetimes having multiple NUT parameters in two simple ways that allow one to circumvent the strong restrictions imposed by the vacuum field equations. First, motivated by momentum relaxation holographic models, we construct multi-NUT spaces in AdS with flat horizons by adding free scalar fields possessing an axionic profile. In our second approach, we build similar configurations in Einstein gravity with quadratic-curvature corrections. We end by presenting planar versions of the Kaluza-Klein monopole in AdS with different magnetic charges. As a byproduct, our strategy yields a simple way to generalize the Gross-Perry-Sorkin monopole to include the cosmological constant.
\end{abstract}

\maketitle

\section{Introduction}

The AdS/CFT correspondence~\cite{Maldacena:1997re,Gubser:1998bc,Witten:1998qj,Witten:1998zw} is a powerful tool for studying strongly coupled systems in condensed matter physics~\cite{Hartnoll:2011fn}. In particular, dyonic AdS black holes with planar horizons have been used to describe holographic superconductors, in which DC conductivities can be obtained from linear perturbations of bulk gauge fields~\cite {Hartnoll:2008kx,Horowitz:2010gk,Horowitz:2009ij,Gregory:2009fj}. In the fluid/gravity correspondence~\cite{Hubeny:2011}, on the other hand, vorticity alone leads to very rich structures in non-dissipative holographic fluids~\cite{Caldarelli:2011idw,Leigh:2011au,Leigh:2012jv,Eling:2013sna,Mukhopadhyay:2013gja}. However, for a holographic fluid to exhibit swirling motion, the dual bulk geometry must have a rotation or NUT parameter turned on. Since vorticity plays a key role in superconductivity and superfluidity, turning on angular momentum on planar black holes could lead to interesting holographic models. 

Despite many advances in describing the twisting motion of holographic fluids, their gravitational duals become much more involved in higher dimensions. This is mainly because increasing the spatial dimensions introduces additional independent rotation planes, leading to multiple angular-momentum charges. The renowned Myers-Perry black holes~\cite{Myers:1986un} could be used in these explorations or configurations with multiple NUT parameters~\cite{Chen:2006ea,Chen:2006xh}. Alternatively, for plane-symmetric black holes, one could rely on the solutions of Ref.~\cite{Awad:2002cz} which possess multiple rotation parameters and, quite remarkably, are amenable to electromagnetic charging. However, for planar black holes, no extensions with multiple NUT parameters (henceforth multi-NUT) exist in Einstein-AdS gravity due to obstructions posed by the vacuum equations~\cite{Mann:2003zh,Mann:2005ra}. For a general multi-NUT metric ansatz, the Einstein field equations are solved as long as simple constraints are satisfied by the NUT parameters, the curvature of their corresponding transverse spaces, and the cosmological constant. When the spacetime has plane symmetry, all the curvatures of the transverse sections vanish, collapsing the constraints into one of two options: either the NUT parameters are all equal, or the cosmological constant must disappear.

A similar difficulty is faced when extending the Kaluza-Klein monopole found by Gross, Perry, and Sorkin (GPS) to AdS space within the four-dimensional Einstein-Maxwell-dilaton theory~\cite{Gross:1983hb,Sorkin:1983ns}. For instance, Onemli and Tekin showed that there is no five-dimensional static Kaluza-Klein monopole in AdS that reduces to its flat counterpart in the vanishing cosmological constant limit~\cite{Onemli:2003gg}. Later, Mann and Stelea found that, indeed, the solution can be extended to include the cosmological constant if it satisfies a relation with the NUT charge~\cite{Mann:2005gk}. However, one cannot turn off the former without setting the latter to zero, obstructing its flat limit. Even more, since in higher dimensions multiple NUT parameters can be considered to construct Kaluza-Klein multi-monopoles, the field equations imply that either the cosmological constant must vanish or the NUT charges must be equal~\cite{Mann:2005gk}. Thus, no planar Kaluza-Klein multi-monopoles are known to exist in the presence of the cosmological constant. 

Another question worth considering is whether generalizing the vacuum equations relaxes the constraints  on the system. It is well known that higher-curvature corrections are inevitable in general relativity if it is treated as an effective field theory. The low-energy limit of string theory yields ghost-free, nontrivial gravitational self-interactions via the Gauss-Bonnet term in dimensions higher than four~\cite{Zwiebach:1985uq,Deser:1986xr}. On the other hand, generic quadratic-curvature corrections in four dimensions are known to improve the ultraviolet behavior of the graviton's propagator, rendering the theory renormalizable at the one-loop level around the Minkowski background, at the price of introducing ghosts~\cite{Stelle:1976gc}. In lower dimensions, the presence of higher-curvature corrections to general relativity endows the particle spectrum with local degrees of freedom~\cite{Bergshoeff:2009hq,Bergshoeff:2009aq}, generating interesting black hole solutions which are not necessarily of constant curvature~\cite{Clement:2009gq,Clement:2009ka,Oliva:2009ip}, some of them being relevant for describing holographic non-relativistic strongly-coupled systems~\cite{Ayon-Beato:2009rgu}. 

In this paper, we study two frameworks that circumvent the aforementioned obstructions. The first one is motivated by a holographic model for momentum relaxation that can be investigated using spatially dependent scalar fields with shift symmetry~\cite{Andrade:2013gsa}. In that case, the bulk gravitational theory is described by the Einstein-Maxwell action supplemented by free scalar fields with axionic profiles. Since stationarity is induced by multiple NUT parameters, the problem becomes difficult to address in the presence of electromagnetic fields. Thus, for simplicity, we consider only the contributions of the axionic fields, leaving the problem of charging multi-NUT spacetimes open. However, by introducing free scalar fields into the fold, our constructions constitute the first asymptotically locally AdS multi-NUT solutions with a planar horizon reported in the literature. They generalize the static solutions of Ref.~\cite{Bardoux:2012aw} and represent the initial steps towards new momentum relaxation models.

Our second construction is motivated by higher-curvature corrections that allow one to introduce an asymptotic anisotropic scaling symmetry~\cite{Son:2008ye,Balasubramanian:2008dm,Kachru:2008yh}. Particular attention is paid to the curve in parametric space where the higher-dimensional Lifshitz black hole was found~\cite{Ayon-Beato:2010vyw}. Here, however, we restrict ourselves to isotropic scaling symmetry; as such, our multi-NUT solutions are generalizations only of the AdS black holes of Ref.~\cite{Ayon-Beato:2010vyw}. The multi-NUT spacetimes in this theory are the first vacuum solutions to possess both plane symmetry and locally AdS asymptotics. Since the equations of motion are now considerably less restrictive, we are also able to build other multi-NUT spacetimes with a number of different horizon geometries beyond those permitted by standard gravity.

The solutions we report here represent the starting point to obtain planar Kaluza-Klein (multi-)monopoles with a nonvanishing cosmological constant. To go further in this direction, first we use the mechanism proposed in Ref.~\cite{Cisterna:2017qrb} to endow the GPS monopole with a cosmological constant with a well-defined $\Lambda\to0$ limit. Then, we construct a multi-NUT-AdS string in odd dimensions by adding a single flat direction to its lower-dimensional counterpart. Following the GPS prescription~\cite{Gross:1983hb,Sorkin:1983ns}, we perform a dimensional reduction along the periodic direction that generates the $U(1)$ fibration over the planar Einstein-K\"ahler manifold in the seed metric. This procedure yields a planar Kaluza-Klein multi-monopole solution to Einstein-Maxwell-dilaton gravity with a Liouville potential and free axionic fields, including a cosmological constant. 

The manuscript is organized as follows: In Sec.~\ref{sec:No-go}, we present the problem setup and explicitly review the obstruction that arises when constructing planar AdS multi-NUT spacetimes in vacuum General Relativity. In Sec.~\ref{sec:axions}, we study a simple framework in which the obstruction can be circumvented by introducing free complex scalar fields with an axionic profile, thereby obtaining the first planar AdS multi-NUT configuration. In addition, the solution's physical properties are discussed. In Sec.~\ref{sec:higher-curvature}, we show that quadratic-curvature corrections to Einstein-AdS gravity allow one to construct multi-NUT as well, without introducing free complex scalar fields. Then, Sec.~\ref{sec:KKmonopoles} is devoted to presenting the planar Kaluza-Klein (multi-)monopoles with nonvanishing cosmological constant. Finally, in Sec.~\ref {sec:conclusions}, we present our conclusions. 

\section{A no-go result in vacuum Einstein-AdS gravity\label{sec:No-go}}

To show that General Relativity does not support planar AdS spacetimes with multiple NUT parameters in arbitrary $D$ dimensions, we start from the Einstein field equations in vacuum, that is,  
\begin{equation} \label{EFE}
     G_{\mu\nu} + \Lambda g_{\mu\nu} = 0\,, 
\end{equation}
where $G_{\mu\nu}=R_{\mu\nu} - \frac{1}{2}g_{\mu\nu} R$ is the Einstein tensor. The plane-symmetric analog of the Schwarzschild-Tangherlini solution can be constructed out of the ansatz for the line element~\cite{Tangherlini:1963bw}
\begin{subequations} \label{blackhole}
\begin{equation}
    \rd s^2 = -f(r)\rd t^2 + \frac{1}{f(r)}\rd r^2 + r^2\rd \Sigma^2\,,
\end{equation}    
with $\rd \Sigma^2$ denoting a flat codimension-2 space, which is locally $\mathbb{R}^{D-2}$. The metric function that solves the field equations~\eqref{EFE} is 
\begin{equation}
    f(r)=-\frac{2\Lambda r^2}{(D-1)(D-2)}-\frac{m}{r^{D-3}}\,,
\end{equation}
\end{subequations}
where $m$ is an integration constant associated with the mass. Notice that this one-parameter family of solutions represents a black hole only for a negative cosmological constant, which is related to the AdS radius via
\begin{align}\label{LambdaEll}
    \Lambda = - \frac{(D-1)(D-2)}{2\ell^2}\,.
\end{align}

Planar black holes can be simultaneously charged and set into general rotation~\cite{Awad:2002cz}, including, of course, the solutions in $D=3,4$~\cite{Banados:1992wn,Clement:1993kc,Lemos:1994xp,Lemos:1995cm}. The latter renders these solutions special when compared to their spherical counterparts. Currently, it remains an open problem to extend the vacuum rotating Myers-Perry black holes~\cite{Myers:1986un} to Einstein-Maxwell theory while retaining full general rotation, although some advances have been made in the presence of scalar fields~\cite{Barrientos:2025abs}. A key advantage that occurs for planar horizons is that the static and rotating solutions are related by large coordinate transformations that, although they keep local properties invariant, change the global structure in a nontrivial way --- see, for instance, Ref.~\cite{Martinez:1999qi}.

In an attempt to add multiple NUT parameters to these black holes, one considers the following ansatz for the line element~\cite{Mann:2003zh,Mann:2005ra}
\begin{subequations} \label{ansatz}
\begin{align}\label{metricmultinut}
    \diff{s^2} &= -f(r)\bigg(\diff{t} + 2\sum_{i=1}^{k}n_{(i)} B_{(i)} \bigg)^2+\frac{\diff{r^2}}{f(r)} + \sum_{i=1}^{k}(r^2+n_{(i)}^2)\diff{\Sigma_{(i)}^2} \,, 
\end{align}
where $k=(D-2)/2$ and $D$ is henceforth taken to be even.\footnote{An odd-dimensional multi-NUT ansatz requires only adding the term $r^2\diff{y}^2$ to Eq.~\eqref{metricmultinut} with $y$ being the additional coordinate in the $(D+1)-$dimensional space~\cite{Mann:2003zh,Mann:2005ra}.} This extension of the black hole metric in Eq.~\eqref{blackhole} uses the direct product structure of the flat space for the base manifold, i.e. $\mathbb{R}^{2k}=\mathbb{R}^2\times \mathbb{R}^2\times\dots\times \mathbb{R}^2$. Consequently, the isometry group of the horizon geometry breaks into the corresponding product of subgroups. However, the ansatz in Eq.~\eqref{metricmultinut} is still highly symmetric as it draws from the K\"ahler geometry of the planes for its construction. The flat plane metrics, $\diff{\Sigma_{(i)}^2}$, are Einstein-K\"ahler manifolds with associated symplectic forms $\Omega_{(i)} = \rd B_{(i)}$. Taking further advantage of their K\"ahler structure, we write them in terms of holomorphic and anti-holomorphic coordinates, i.e. $z_{(i)} = x_{(i)} + \ri \,y_{(i)}$ and $\bar{z}_{(i)}=x_{(i)} - \ri \,y_{(i)}$, respectively, as follows
    \begin{align}\label{dSigmai}
    \diff{\Sigma}_{(i)}^2 &= \frac{1}{2}\left(\diff{\bar{z}_{(i)}}\diff{z_{(i)}}+\diff{z_{(i)}}\diff{\bar{z}_{(i)}}\right),\\
    B_{(i)} &= \frac{1}{4 \ri}\left(\bar{z}_{(i)}\diff{z_{(i)}}-z_{(i)}\diff{\bar{z}_{(i)}}\right),\label{Bi}
    \end{align}
\end{subequations}
where the sum over $(i)$ is not assumed unless otherwise stated. From this construction, one can obtain the nonvanishing components of the Ricci tensor; they are
\begin{subequations}\label{Riccicomp}
    \begin{align}
        R^t_t = R^r_r &= -\frac{1}{2} f'' - \sum_{i=1}^k \frac{r f'}{r^2+n_{(i)}^2} - \sum_{i=1}^k \frac{2 n_{(i)}^2 f}{(r^2+n_{(i)}^2)^2}\,, \\
        R^{x_{(i)}}_{x_{(i)}} = R^{y_{(i)}}_{y_{(i)}} &= -\frac{rf'}{r^2+n_{(i)}^2} - \frac{(r^2-n_{(i)}^2)}{(r^2+n_{(i)}^2)^2}\,f - \frac{2r^2f}{r^2+n_{(i)}^2}\sum_{j\neq i}\frac{1}{r^2+n_{(j)}^2}\,,       
    \end{align}
\end{subequations}
where prime denotes differentiation with respect to the radial coordinate $r$. Then, the Ricci scalar can be computed directly from the diagonal components of the Ricci tensor, giving
\begin{align}\label{Ricciscal}
    R = -f'' - 4 \sum_{i=1}^{k} \frac{r f'}{r^{2} + n_{(i)}^{2}} - 2 f \sum_{i=1}^{k} \frac{1}{r^{2} + n_{(i)}^{2}} - 8 r^{2} f \sum_{i=1}^{k} \sum_{j>i}^{k} \frac{1}{(r^{2} + n_{(i)}^{2})(r^{2} + n_{(j)}^{2})}\,.
\end{align}
The remaining nontrivial off-diagonal component can be read off from the contracted Bianchi identity $\nabla_\mu G^\mu_\nu=0$, giving $R^t_{y_{(i)}} = 2n_{(i)}x_{(i)}[R^t_t-R^{x_{(i)}}_{x_{(i)}}]$, since it is not independent of Eqs.~\eqref{Riccicomp} and~\eqref{Ricciscal} because of diffeomorphism invariance.

Let us now provide a streamlined proof of the no-go result of Ref.~\cite{Mann:2003zh,Mann:2005ra}. To do so, we start by considering the two independent components of the Einstein equations, say, the $tt$ and $x_{(i)}x_{(i)}$ components, which yield
\begin{subequations}\label{EFEvacmulti}
    \begin{align}\label{EFEvacmultitt}
    0 &= \sum_{i=1}^k \frac{r f'}{r^{2} + n_{(i)}^{2}}
    + f \sum_{i=1}^k \frac{r^2 - n_{(i)}^2}{(r^{2} + n_{(i)}^{2})^2}
    + 4 r^{2} f \sum_{i=1}^{k} \sum_{j>i}^{k} \frac{1}{(r^{2} + n_{(i)}^{2})(r^{2} + n_{(j)}^{2})} + \Lambda\,, \\ \notag
    0 &= \frac{1}{2} f''  - \frac{r f'}{r^{2} + n_{(i)}^{2}} - \frac{\left( r^{2} - n_{(i)}^{2} \right)}{(r^{2} + n_{(i)}^{2})^2} f +  \sum_{j=1}^k \frac{2r f' + f}{r^{2} + n_{(j)}^{2}}  -\frac{2r^{2}f}{r^{2} + n_{(i)}^{2}} \sum_{j \ne i}^k \frac{1}{r^{2} + n_{(j)}^{2}} \\ & \quad + 4 r^{2} f \sum_{i=1}^{k} \sum_{j>i}^{k} \frac{1}{(r^{2} + n_{(i)}^{2})(r^{2} + n_{(j)}^{2})} + \Lambda\,. \label{EFEvacmultixx}
\end{align}
\end{subequations}
In order to integrate the Eq.~\eqref{EFEvacmultitt}, we introduce the function 
\begin{align}\label{Pdef}
    P_{(k)}(r) := \prod_{i=1}^k(r^2+n_{(i)}^2)\,, \quad \text{satisfying} \quad \frac{\diff{}}{\diff{r}} \ln P_{(k)}(r) := (\ln P_{(k)})' = 2r\sum_{i=1}^k \frac{1}{r^2+n_{(i)}^2}\,.
\end{align}
Using the identity
\begin{align}
    \left(\sum_{i=1}^k\frac{1}{r^2+n_{(i)}^2}\right)^2 = \sum_{i=1}^k\frac{1}{(r^2+n_{(i)}^2)^2} + 2\sum_{i=1}^{k} \sum_{j>i}^{k} \frac{1}{(r^{2} + n_{(i)}^{2})(r^{2} + n_{(j)}^{2})} \,,
\end{align}
and the definition of $P_{(k)}(r)$ in Eq.~\eqref{Pdef}, the $tt$ components of the Einstein equations, after some algebra, can be rewritten as
\begin{align}
  \frac{1}{2}\left[f' + f\left(\ln\frac{P_{(k)}}{r} \right)'\, \right](\ln P_{(k)})' + \Lambda = 0\,.
\end{align}
Multiplying this equation on both sides by $P_{(k)}/r$, the left-hand side can be written as a total derivative, whose first integral yields 
\begin{align}
    f(r) = -\frac{r}{P_{(k)}(r)}\left(m+2\Lambda\int^r\frac{P^2_{(k)}(\rho)\,\diff{\rho}}{\rho\,P'_{(k)}(\rho)} \right)\,,
\end{align}
where $m$ is an integration constant. Inserting this solution into transverse components of the field equations, one finds that the following constraint must be satisfied
\begin{equation}\label{vacuumconst}
    \Lambda(n_{(i)}^2-n_{(j)}^2)=0\,, \quad \forall\ i,j\,.
\end{equation}

Since the spacetime is stationary and there are no curvature singularities, see the discussion below, the range of the coordinates covering it is $-\infty<t<\infty$, $-\infty<r<\infty$, $-\infty<\Re{(z_{(i)})}<\infty$, and $-\infty<\Im{(z_{(i)})}<\infty$ for all $i$. Most holographic models would consider noncompact transverse spaces, as above. However, cylindrical or toroidal symmetry may be considered at any point. For instance, if rotation parameters are enabled (via a large-gauge transformation~\cite{Awad:2002cz}), an identification must be performed in each rotation plane. Furthermore, one could consider a periodic time coordinate in the Euclidean section or in compactifications of Kaluza-Klein theory, for instance. In particular, in Sec.~\ref{sec:KKmonopoles}, we consider a periodic Euclidean time coordinate.

In other words, Eq.~\eqref{vacuumconst} implies that either the cosmological constant must be set to zero or all NUT parameters must be equal to each other~\cite{Mann:2005ra}. The former case is undesirable, as it avoids asymptotically AdS spaces with multi-NUT parameters, while the latter recovers the higher-dimensional version of the planar Taub-NUT spacetime; such spaces are well-known in the literature~\cite{Bais:1984xb,Page:1985bq,Taylor:1998fd,Awad:2000gg}. Our main purpose is to circumvent this restriction imposed by the Einstein vacuum equations when considering planar transverse sections. In what follows, we analyze two separate scenarios and build explicit examples of plane-symmetric asymptotically AdS multi-NUT spacetimes.

\section{Shaping Multi-NUT Spacetimes with Free Fields\label{sec:axions}}

The first scenario is motivated by the simplest holographic model of momentum relaxation~\cite{Andrade:2013gsa}. In that setup, momentum relaxation is realized by allowing spatial dependence of the holographic sources to the dual operators of massless scalar fields. Crucially, this dependence can be chosen so that the bulk geometry and the energy-momentum tensor of massless scalars are invariant under the same isometry group, although the massless scalar fields are not. A similar strategy has been used to construct planar black holes, homogeneous black strings, and black branes in Einstein-AdS gravity~\cite{Bardoux:2012aw,Cisterna:2017qrb}. 

The setup in Ref.~\cite{Andrade:2013gsa} is Einstein-Maxwell-AdS with $D-2$ free real scalar fields. Since charging multi-NUT spacetimes poses a significant challenge, we focus on the electrically neutral version of that model. However, for convenience, we formulate the matter sector as composed of $k=(D-2)/2$ free massless complex scalar fields $\varphi_{(i)}$ instead, whereas $\bar{\varphi}_{(i)}$ denotes their complex conjugate.\footnote{We use the normalization $\varphi_{(i)} = \Phi_{(i)} + i\,\Psi_{(i)}$, where $\Phi_{(i)}$ and $\Psi_{(i)}$ are real scalar fields.} Concretely, we consider the action principle
\begin{equation}\label{Ibulk}
    I_{\rm bulk} =  \int_{\mathcal{M}} \rd^D x \sqrt{|g|}\left(\frac{R-2\Lambda}{16\pi}-\frac{1}{2}\sum\limits_{i=1}^{k}|\nabla\varphi_{(i)}|^2\right)\,,
\end{equation}
where we are using units such that the Newton constant is $G_N=1$. Arbitrary variations with respect to the metric and the scalar fields lead to
\begin{subequations}\label{eom}
    \begin{align}\label{eomg}
    R_{\mu\nu} - \frac{1}{2}g_{\mu\nu} R + \Lambda g_{\mu\nu} &= 8\pi\,T^{(\varphi)}_{\mu\nu},  \\
        \Box\varphi_{(i)} &=0,
    \end{align}
    respectively, where the energy-momentum tensor of the complex scalar fields is given by
\begin{align} \label{Tmunuvarphi}
    T_{\mu\nu}^{(\varphi)} &= \frac{1}{2}\sum_{i=1}^{k}\Big(\nabla_{\mu}\bar{\varphi}_{(i)}\nabla_{\nu}\varphi_{(i)}+\nabla_{\mu}\varphi_{(i)}\nabla_{\nu}\bar{\varphi}_{(i)}  -g_{\mu\nu} \nabla_{\alpha}\bar{\varphi}_{(i)}\nabla^{\alpha}\varphi_{(i)}\Big) \,.
\end{align}
\end{subequations}

To solve the field equations~\eqref{eom}, the metric ansatz~\eqref{ansatz} is assumed. If one relaxes the condition that the scalar field must remain invariant under the action of the isometry group of the metric, while keeping the stress-energy tensor invariant, one finds that the Klein-Gordon equation is solved by ~\cite{Andrade:2013gsa,Bardoux:2012aw}
\begin{equation}\label{varphisol}
    \varphi_{(i)} = \lambda_{(i)} z_{(i)},
\end{equation}
which is also known to be compatible with Taub-NUT spacetimes in four dimensions~\cite{Bardoux:2013swa}. Notice that, by using complex fields, the global $U(1)$ symmetry of our field content is evident. At the same time, the fields are fully adapted to our metric ansatz~\eqref{ansatz}, as they are linear in the coordinates that span the codimension-2 hypersurface of constant $t$ and $r$. Direct calculation of the independent components of the stress tensor~\eqref{Tmunuvarphi} yields
\begin{subequations}
    \begin{align}
    T^t_t = T^r_r &= - \sum_{i=1}^k\frac{\lambda_{(i)}^2}{r^2+n_{(i)}^2} \,,\\
    T^{x_{(i)}}_{x_{(i)}} = T^{y_{(i)}}_{y_{(i)}} &= \frac{\lambda_{(i)}^2}{r^2+n_{(i)}^2} - \sum_{j=1}^k\frac{\lambda^2_{(j)}}{r^2+n^2_{(j)}}\,.
\end{align}
\end{subequations}
As in the vacuum case, the non-vanishing off-diagonal components are not independent of these ones, as a consequence of diffeomorphism invariance. Then, the independent components of the field equations are 
\begin{subequations}\label{EFEmulti}
    \begin{align}\notag
    0 &= \sum_{i=1}^k \frac{r f'}{r^{2} + n_{(i)}^{2}}
    + f \sum_{i=1}^k \frac{r^2 - n_{(i)}^2}{(r^{2} + n_{(i)}^{2})^2}
    + 4 r^{2} f \sum_{i=1}^{k} \sum_{j>i}^{k} \frac{1}{(r^{2} + n_{(i)}^{2})(r^{2} + n_{(j)}^{2})} \\ &\quad + \Lambda + \sum_{i=1}^k\frac{8\pi\lambda_{(i)}^2}{r^2+n_{(i)}^2}\,, \label{EFEmultitt} \\ \notag
   0 &= \frac{1}{2} f''  - \frac{r f'}{r^{2} + n_{(i)}^{2}} - \frac{\left( r^{2} - n_{(i)}^{2} \right)}{(r^{2} + n_{(i)}^{2})^2} f +  \sum_{j=1}^k \frac{2r f' + f}{r^{2} + n_{(j)}^{2}}  -\frac{2r^{2} f}{r^{2} + n_{(i)}^{2}} \sum_{j \ne i}^k \frac{1}{r^{2} + n_{(j)}^{2}} \\ & \quad + 4 r^{2} f \sum_{i=1}^{k} \sum_{j>i}^{k} \frac{1}{(r^{2} + n_{(i)}^{2})(r^{2} + n_{(j)}^{2})} + \Lambda - \frac{8\pi\lambda_{(i)}^2}{r^2+n_{(i)}^2} + \sum_{j=1}^k\frac{8\pi\lambda^2_{(j)}}{r^2+n^2_{(j)}} \,. \label{EFEmultixx}
\end{align}
\end{subequations}
Using the same identities as in the vacuum case, Eq.~\eqref{EFEmultitt} can be written as 
\begin{align}
  \frac{1}{2}\left[f' + f\left(\ln\frac{P_{(k)}}{r} \right)'\, \right](\ln P_{(k)})' + \Lambda +  \sum_{i=1}^k\frac{8\pi\lambda_{(i)}^2}{r^2+n_{(i)}^2} = 0\,,
\end{align}
which can be integrated, giving
\begin{subequations}\label{solution1}
\begin{align}\label{f(r)_1}
f(r) = -\frac{r}{P_{(k)}(r)}\left(m+2\int^r\diff{\rho}\left[\Lambda  + \sum_{i=1}^k\frac{8\pi\lambda_{(i)}^2}{\rho^2+n_{(i)}^2}\right]\frac{ P^2_{(k)}(\rho)}{\rho\,P'_{(k)}(\rho)}\right)\,.
\end{align}
Inserting this solution into the transverse components, one finds that the vacuum constraint~\eqref{vacuumconst} is modified according to
\begin{equation}\label{lambdaconstraint}
    \lambda_{(i)}^2 =\lambda_{(k)}^2 + \frac{\Lambda}{8\pi k} \left(n_{(k)}^2-n_{(i)}^2\right)\,, \quad \forall\, i\;\; \mbox{with $k$ fixed}.
\end{equation}
In other words, by fixing all but one of the field's integration constants, we have constructed multi-NUT spacetimes with the highest number of NUT parameters allowed by the dimension. Notice that the metric function in Eq.~\eqref{f(r)_1} is similar to that in Ref.~\cite{Mann:2005ra} [cf. Eq.~\eqref{Pdef}], but it includes the backreaction of the axionic fields that gravitate as an effective negative curvature~\cite{Bardoux:2012tr}. Additionally, due to the presence of $\lambda_{(i)}$, the constraint in Eq.~\eqref{lambdaconstraint} allows one to relax the strong conditions placed on either the cosmological constant or the different NUT charges found in~\cite{Mann:2005ra}. Furthermore, notice that this solution has a smooth static limit when $n_{(i)}\to0$, recovering the well-known solutions of Ref.~\cite{Bardoux:2012aw} with a nonvanishing cosmological constant. Even more, the solutions in Ref.~\cite{Mann:2005ra} are recovered in the vacuum limit, i.e. when $\lambda_{(i)}\to0,\,\forall i$. 
Lastly, reality conditions on the scalar fields demand that the right-hand side of Eq.~\eqref{lambdaconstraint} be positive definite. Then, $\lambda_{(k)}$ must satisfy the bound
\begin{align}
    \lambda_{(k)}^2 > -\frac{\Lambda}{8\pi k} \left(n_{(k)}^2-n_{(i)}^2\right)\,.
\end{align}
Without loss of generality, we assume that $n_{(k)}>n_{(i)}$. Hence, the parameter ranges are restricted whenever the cosmological constant is negative, which is the case we focus on from here on. On the other hand, for a positive cosmological constant, any $\lambda_{(k)}\in\mathbb{R}$ satisfies the bound, and there is no restriction whatsoever on its value.
\end{subequations}

The solution is free of curvature singularities, as it can be checked by analyzing the behavior of its invariants $\forall\, r\in\mathbb{R}$ directly. Additionally, although the complex scalar fields are linear in the coordinates that parametrize the flat transverse sections, one can check that all invariants constructed out of their stress tensor are also regular. Even more, the solution is devoid of Misner strings, similar to the planar Taub-NUT-AdS solution in four-dimensional general relativity~\cite{Misner:1963fr,Chamblin:1998pz,Astefanesei:2004kn}. To see this, let us recall Misner's original analysis~\cite{Misner:1963fr}. Upon calculating $(\nabla t)^2$ for the spherical Taub-NUT solution, he found that it diverges along an infinite line, say, at the south pole. A group-theoretic analysis then shows that this is only a coordinate singularity, provided the time coordinate is periodic. It follows that, for the spherical Taub-NUT geometry, one can either choose spacetime topologies that carry a line defect, such as a cosmic string, or those with closed timelike curves. For the planar Taub-NUT spacetime $(\nabla t)^2$ does not become infinite anywhere, signaling the absence of line defects such as those found in Refs.~\cite{Chamblin:1998pz,Astefanesei:2004kn}. Yet, for reasons unrelated to Misner strings, these solutions possess closed timelike curves~\cite{Astefanesei:2004kn}.

For our multi-NUT spacetimes, it can be verified that
\begin{equation}
    (\nabla t)^2 = -\frac{1}{f(r)}+\sum\limits_{i=1}^{k} \frac{n_{(i)}^2|z_{(i)}|^2}{r^2+n_{(i)}^2},
\end{equation}
does not become infinite for any value of the transverse coordinates $z_{(i)}$, independently of any identification taken in the base space. Consequently, $t$ has no required periodicity. If, however, a periodicity is imposed on $t$ then the spacetime presents quasiregular singularities~\cite{Konkowski:1985ugk,*Konkowski:1985tq}. These mild singularities are the endpoints of incomplete and inextensible geodesics that spiral infinitely around a topologically closed spatial dimension. For any observer approaching the singularity, their world line ends in a finite proper time, without encountering infinite tidal forces. Naturally, forcing the $t$ coordinate to be periodic means that there are closed timelike curves in the regions of spacetime where it is timelike. Nevertheless, for any spacetime topology admitted by planar multi-NUT systems there are always closed timelike curves that extend to infinity on the transverse space, just as in the single NUT parameter case~\cite{Astefanesei:2004kn}.

For a negative cosmological constant given by Eq.~\eqref{LambdaEll}, the planar multi-NUT solution presented here is asymptotically locally AdS, as can be checked by inspecting the behavior of the Riemann curvature near the conformal boundary. Indeed, asymptotically, the metric function behaves as
\begin{align}
    f(r) = \frac{r^2}{\ell^2} -\frac{1}{2k-1}\left(8\pi\lambda_{(k)}^2 + \frac{(2k+3)n_{(k)}^2}{\ell^2} + \frac{2}{\ell^2}\sum_{i=1}^{k-1}n_{(i)}^2 \right)  + \mathcal{O}(r^{-2})\,,
\end{align}
as $r\to\infty$. Thus, it is clear that axions and the multi-NUT parameters gravitate as an effective transverse-curvature section, although the latter is flat. 

To obtain the renormalized Euclidean on-shell action, the bulk functional in Eq.~\eqref{Ibulk} must be supplemented by the Gibbons-Hawking-York term and intrinsic boundary counterterms that render the variational principle for Dirichlet boundary conditions of the holographic sources well posed~\cite{Balasubramanian:1999re,Emparan:1999pm,deHaro:2000vlm,Bianchi:2001kw,Skenderis:2002wp}. In the case of Einstein-dilaton-AdS gravity coupled with axions, this was studied in four dimensions in Ref.~\cite{Caldarelli:2016nni} (see also Ref.~\cite{Anastasiou:2024xhr}). Concretely, in $D=d+1$ dimensions, the full action is $I=I_{\rm bulk} + I_{\rm GHY} + I_{\rm ct}$, where
\begin{align}
     I_{\rm GHY} &= \frac{1}{8\pi}\int_{\partial\mathcal{M}}\diff{^{d}x}\sqrt{|h|}\, K\,, 
\end{align}
is the Gibbons-Hawking-York term, with $h=\det h_{ij}$ being the induced metric on $\partial\mathcal{M}$ and $K=h^{ij}K_{ij}$ the trace of the extrinsic curvature. Additionally, the intrinsic boundary counterterms are given by~\cite{Balasubramanian:1999re,Emparan:1999pm,deHaro:2000vlm,Bianchi:2001kw,Skenderis:2002wp}
\begin{align}
    \notag
    I_{\rm ct}  = \frac{1}{8\pi}\int_{\partial\mathcal{M}}\diff{^{d}x}\sqrt{|h|}\Bigg[& \frac{d-1}{\ell} + \frac{\ell}{2(d-2)}\mathcal{R} + \frac{\ell^3}{2(d-4)(d-2)^2}\left(\mathcal{R}_{ij}\mathcal{R}^{ij} - \frac{d}{4(d-1)}\mathcal{R}^2 \right) \notag \\ &\qquad + \frac{\ell}{2(2\Delta - d-2)}\sum_{i=1}^{k}h^{m n}\mathcal{D}_m\bar{\varphi}_{(i)}\mathcal{D}_n\varphi_{(i)} + ... \Bigg] \label{Ict}
\end{align}
with $\Delta$ being the conformal weight of the scalar fields, while ellipsis denotes higher-derivative terms in the metric and complex scalars. Since we are interested in the massless case, we focus on the case $\Delta=d$, where the scalar operator in the dual CFT is marginal. 

For simplicity, we focus on the simplest nontrivial AdS multi-NUT solution, namely, the case $k=2$ ($d=5$). The analytic continuation of the line element~\eqref{metricmultinut} with metric function~\eqref{solution1} can be obtained by performing the Wick rotation $t\to-\ri\tau$ and $n_{(i)}\to -\ri n_{(i)}$. The absence of conical singularities at $r=r_h$, defined as the largest positive root of $f(r_h)=0$, demands that $\tau\sim\tau+\beta_\tau$, where 
\begin{align}\label{beta}
    \beta_\tau = \frac{4\pi}{f'(r)}\bigg|_{r=r_h} = \frac{4\pi\ell^2 r_h}{5(r_h^2-n_{(2)}^2) - 8\pi\ell^2\lambda_2^2}\,.
\end{align}
From here on, we assume, without loss of generality, that $r_h>n_2>n_1$. The counterterms in Eq.~\eqref{Ict} renormalize all but one divergence of the bulk action plus the Gibbons-Hawking-York term: the linear divergence on the radial coordinate. This is an indication that a local counterterm for the scalar field is missing. To systematically renormalize the theory, one needs to perform holographic renormalization~\cite{deHaro:2000vlm,Bianchi:2001kw,Skenderis:2002wp} in the presence of axionic fields in six dimensions, similar to what was done in Ref.~\cite{Caldarelli:2016nni} in the four-dimensional case. One would expect higher-derivative terms for the axionic fields from this analysis. However, this lies beyond the scope of this paper, since our aim is to present a mechanism to circumvent obstructions imposed by the field equations to the existence of planar multi-NUT solutions in AdS, rather than to renormalizing Einstein-AdS coupled to axionic fields in six dimensions. Nevertheless, a suitable choice of parameters can eliminate the remaining divergence and render the action completely finite. This condition fixes the remaining axionic constant in terms of the two free NUT charges, according to 
\begin{align}
    \lambda_{(2)}^4 &- \frac{\left(4n_{(1)}^{2} - n_{(2)}^2\right)\lambda_{(2)}^2}{4\pi \ell^2} + \frac{5\left(n_{(1)}^2-n_{(2)}^2\right)\left(7n_{(1)}^2+5n_{(2)}^2\right)}{256\pi^2 \ell^4} = 0\,.\label{constlambda2}
\end{align}
This condition is consistent with the planar Taub-NUT-AdS solution without axionic fields reported in Refs.~\cite{Bais:1984xb,Page:1985bq,Taylor:1998fd,Awad:2000gg} when $n_{(1)}=n_{(2)}=n$, since this constraint alongside Eq.~\eqref{lambdaconstraint} implies that both axions vanish. Then, the renormalized Euclidean on-shell action for the six-dimensional solution with two independent NUT charges is given by
\begin{align} \label{FreeEnergy}
    I_E = \frac{\beta_{\tau}\Sigma}{16\pi}\bigg(m - 4\pi r_h(n_{(1)}^2-n_{(2)}^2)(\lambda_{(1)}^2-\lambda_{(2)}^2)  - \frac{r_h\left[12r_h^4 - 20r_h^2(n_{(1)}^2+n_{(2)}^2) + 15(n_{(1)}^2+n_{(2)}^2)^2 \right]}{6\ell^2}\bigg)\, ,
\end{align}
where $\Sigma$ denotes the volume of the codimension-2 transverse section of constant $t-r$, and $I_E$ is defined with an overall minus sign with respect to its Lorentzian counterpart, as usual.

The mass of the six-dimensional planar AdS multi-NUT configuration can be obtained by following the Hamiltonian formulation of the theory in Eq.~\eqref{Ibulk}. For simplicity, we will consider the matter sector in terms of $2k$ real scalar fields, $\phi_{(i)}$, instead of the $k$ free complex ones, $\varphi_{i}$. Thus, the action takes the following equivalent form,
\begin{equation}
I_{\mathrm{bulk}}=\int_{\mathcal{M}} \mathrm{d}^D x \sqrt{|g|}\left(\frac{R-2 \Lambda}{16 \pi}-\frac{1}{2} \sum_{i=1}^{2k}\nabla^{\mu} \phi_{(i)}\nabla_{\mu} \phi_{(i)}\right)
\end{equation}
This formulation requires decomposing spacetime into space and time. For this purpose, we apply the ADM decomposition to the Lorentzian metric of our solution,
\begin{equation}
\diff{s^2}=-(N \diff{t})^2+\gamma_{a b}\left(\diff{} x^a+N^a \diff{} t\right)\left(\diff{} x^b+N^b \diff{} t\right)\ ,
\end{equation}
where $\gamma_{ab}$ is the induced metric of a spacelike codimension-1 hypersurface, $\Sigma_t$, taken at constant time $t$, with a unit normal vector $n^a$. The canonical momenta conjugate to the spatial metric $\gamma_{a b}$ and the scalar fields $\phi_{(i)}$ are,
\begin{align}
\pi^{ab}=-\frac{\sqrt{\gamma}}{16\pi}(K^{ab}-\gamma^{ab}K)\ ,\qquad \mbox{and} \qquad 
\pi_{\phi_{(i)}}=-\frac{\sqrt{\gamma}}{N}N^a\nabla_{a}\phi_{(i)}\ ,
\end{align}
respectively, where $K_{ab}=-\nabla_b n_a$ is the extrinsic curvature of $\Sigma_t$, and the operator $\nabla_a$ stands for the covariant derivative compatible with $\gamma_{ab}$. We have also specialized the canonical momentum $\pi_{\phi_{(i)}}$ to a stationary scalar field by using $\dot{\phi}=0$ in the process. Next, we follow the Hamiltonian formalism described in Ref.~\cite{ReggeTeitelboim} for the Einstein–Hilbert action, now including the contributions of the free scalar fields. The resulting Hamiltonian is 
\begin{equation}
H=\int_{\Sigma_t}\mathrm{d}^d x\sqrt{\gamma}(N \mathscr{H}+N^a\mathscr{H}_a)+E\ ,
\end{equation}
where $\gamma=\det\gamma_{ab}$ is the determinant of the induced metric on $\Sigma_t$, and the Hamiltonian constraints are
\begin{align}\notag 
& \mathscr{H}=\frac{16\pi}{\sqrt{\gamma}}\left(\pi_{a b} \pi^{a b}-\frac{1}{D-2} \pi^2\right)-\sqrt{\gamma} \left(\frac{R-2\Lambda}{16\pi}\right)+\frac{1}{2}\sum_{i=1}^{D-2}\left(\frac{\pi_{\phi_{(i)}}^2}{\sqrt{\gamma}}+\sqrt{\gamma}\gamma^{ab}\nabla_a\phi_{(i)}\nabla_b\phi_{(i)}\right)\ ,\\
& \mathscr{H}_a=-2 \nabla_b\pi_a{ }^b+\sum_{i=1}^{D-2}\pi_{\phi_{(i)}}\nabla_a\phi_{(i)}.
\end{align}
The supplementary quantity $E$ ensures a well-defined variational principle, i.e., its variation cancels out all the contributions coming from variations of the bulk action. Concretely, its variation is given by
\begin{equation}\label{deltaE}
\begin{aligned}
\delta E=& \int_B \mathrm{d}^{d-1}\mathrm{s}_m \Big\{\frac{1}{16\pi}G^{ablm}\!\left(N\nabla_l\delta\gamma_{ab}-\partial_l N\,\delta\gamma_{ab}\right)+\left[2N_l\,\delta\pi^{lm}
        + \left(2N^l\pi^{bm}-N^m\pi^{bl}\right)\delta\gamma_{bl}\right] \\
&- \sum_{i=1}^{D-2}\left(\sqrt{\gamma}\,N\,\gamma^{ml}\nabla_l\phi_{(i)}
        + N^m\pi_{\phi_{(i)}}\right)\delta\phi_{(i)}
\Big\}.
\end{aligned}
\end{equation}
with,
\begin{equation}
G^{ablm}=\frac{1}{2}\sqrt{\gamma}\left(\gamma^{al}\gamma^{bm}
+\gamma^{am}\gamma^{bl}-2\gamma^{ab}\gamma^{lm}\right)\ ,
\end{equation}
and it is evaluated on a codimension-2 hypersurface $B$ which is the boundary of $\Sigma_t$.

The quantity $E$ is associated with the mass of the solution when $B$ is evaluated at $r\to\infty$, since it corresponds to the value of the on-shell Hamiltonian. Using the asymptotic behavior of our six-dimensional solution with $k=2$ ($d=5$), Eq.~\eqref{deltaE} can be integrated to obtain $E$. Thus, the mass is given by 
\begin{equation}
E= \frac{m\Sigma}{4\pi}\,,
\end{equation}
where $\Sigma$ is the volume of the codimension-2 hypersurface of constant $t-r$. It is worth noticing that, in the integration of Eq.~\eqref{deltaE}, potential divergences arising from terms with the cosmological constant, which are absent with flat asymptotics, are canceled out by the relation \eqref{lambdaconstraint}. Therefore, this relation ensures the integrability and finiteness of the energy for planar AdS multi-NUT spacetimes.

\section{Higher-curvature corrections to multi-NUT spacetimes\label{sec:higher-curvature}}

In the previous section, matter fields were added to the vacuum Einstein field equations to allow for multiple NUT parameters. Here, we instead consider a scenario in which the vacuum equations receive higher-curvature corrections. In particular, we focus on a quadratic term in the Ricci scalar, whose dynamics are dictated by the action principle
\begin{equation} \label{GR+R2}
    I = \int\rd^Dx\sqrt{|g|} \left( R-2\Lambda +\beta R^2 \right):=\int\rd^Dx\sqrt{|g|}\,\Lag_{R^2} \,,
\end{equation}
where $\beta$ is a coupling constant of mass dimension equal to minus two. Although this one-parameter family of higher-curvature corrections is fourth order in derivatives, it can be mapped into a scalar-tensor theory by performing field redefinitions and conformal transformations --- for details, see Refs.~\cite{Barrow:1988xh,Teyssandier:1983zz,Wands:1993uu,Flanagan:2004bz,Sotiriou:2006hs,Sotiriou:2008rp,DeFelice:2010aj} and the references therein. 

The field equations are obtained by performing arbitrary variations of the action~\eqref{GR+R2} with respect to the metric, giving
\begin{equation}
    R_{\mu\nu} - \frac{1}{2}g_{\mu\nu} R + \Lambda g_{\mu\nu} +\beta H_{\mu\nu}=0\,,\label{eomR2}
\end{equation}
where the higher-derivative terms are encoded in the last tensor, defined as
\begin{equation}
    H_{\mu\nu} = 2g_{\mu\nu}\Box R -2\nabla_{\mu}\nabla_{\nu}R +2RR_{\mu\nu}-\frac{1}{2}R^2g_{\mu\nu}\,.
\end{equation}

In generic higher-curvature theories, vacuum states characterized by maximally symmetric spaces have an effective curvature radius $\ell_{\rm eff}$. In the case of action~\eqref{GR+R2}, these are characterized by the equation
\begin{align}\label{eqLambdaeff}
    2D\beta(D-4)\,\Lambda_{\rm eff}^2 = (D-2)^2\left(\Lambda - \Lambda_{\rm eff} \right) \,, \quad \mbox{where} \quad \Lambda_{\rm eff} = -\frac{(D-1)(D-2)}{2\ell^2_{\rm eff}}\,.
\end{align}
Although $\Lambda_{\rm eff}=\Lambda$ in four dimensions, this is no longer true in higher dimensions, as the presence of $\beta$ modifies the definition of the effective curvature radius. 

This theory is known to allow for planar AdS black holes and asymptotically Lifshitz black holes in higher dimensions at the particular curve in parameter space~\cite{Ayon-Beato:2010vyw}
\begin{equation} \label{R2beta}
    \beta = -\frac{1}{8\Lambda}\,.
\end{equation} 
Along this curve, the discriminant of the quadratic equation for $\Lambda_{\rm eff}$ in~\eqref{eqLambdaeff} becomes $\Delta=4$ in arbitrary dimensions, being independent of $\Lambda$, $\beta$, and $D$. Furthermore, the conformal mapping that brings the action~\eqref{GR+R2} into a scalar-tensor theory cannot be performed since the conformal transformation is degenerate there. Indeed, as pointed out in Ref.~\cite{Ayon-Beato:2010vyw}, this is true for every theory whose Lagrangian can be written as $\Lag=(R-4\Lambda)^2H(R)$ for some arbitrary function $H(R)$ that is regular at $R=4\Lambda$. The present case is the simplest for which the function is constant, that is, $H=-1/(8\Lambda)$. Therefore, the theory~\eqref{GR+R2} is purely gravitational at the point~\eqref{R2beta}, in the sense that it cannot be mapped into a scalar-tensor analog. We will consider this case from here on.

The condition $R=4\Lambda$ not only represents a stationary point of the action, but it solves the field equations automatically. This can be seen by noticing that, if the condition~\eqref{R2beta} is satisfied, the field equations~\eqref{eomR2} can be written as
\begin{align}
    E_{\mu}{}^{\lambda\rho\sigma}R_{\nu\lambda\rho\sigma} + \frac{1}{16\Lambda}(R-4\Lambda)^2 g_{\mu\nu} - 2\nabla^\lambda\nabla^\rho E_{\mu\lambda\rho\nu}  = 0\,,
\end{align}
where $E^{\mu\nu}_{\lambda\rho}$ is the functional derivative of the quadratic Lagrangian~\eqref{GR+R2}, i.e.,
\begin{align}\label{E-tensor}
    E^{\mu\nu}_{\lambda\rho} := \frac{\partial\Lag_{R^2}}{\partial R^{\lambda\rho}_{\mu\nu}} \Bigg|_{\beta=-\frac{1}{8\Lambda}} = -\frac{1}{4\Lambda}(R-4\Lambda)\,\delta^{[\mu}_{[\lambda}\delta^{\nu]}_{\rho]}\,.
\end{align}
Therefore, integrating the second-order condition $R=4\Lambda$ suffices to solve the whole system, without further restrictions on either the cosmological constant or the NUT charges. Assuming the multi-NUT ansatz~\eqref{ansatz}, whose Ricci scalar is given in Eq.~\eqref{Ricciscal}, the latter condition can be rewritten as
\begin{align}
     f'' + 2\frac{P'_{(k)}}{P_{(k)}} f' + \frac{P_{(k)}''}{P_{(k)}} f + 4\Lambda = \frac{(P_{(k)}f)''}{P_{(k)}} + 4\Lambda = 0,
\end{align}
with $P_{(k)}$ defined in Eq.~\eqref{Pdef}. Integrating this equation twice, we obtain the analytic solution of the system, that is,
\begin{align}
f(r) = -\frac{r}{P_{(k)}(r)}\left(m-\frac{b}{r} + \frac{4\Lambda}{r}\int^r\diff{u}\int^u\diff{\rho}\,P_{(k)}(\rho) \right)  \,.
\end{align}
where $m$ and $b$ are integration constants. Notice that only two independent integration constants appear in the solution despite the fact that the model has fourth-order equations. This is a consequence of the fact that the nonlinear equations that determine the metric function $f(r)$ can be solved by the second-order equation $R=4\Lambda$ along the curve in parameter space given by Eq.~\eqref{R2beta}. Such is also the case in the limit where every NUT parameter vanishes, for which the AdS black holes of Ref.~\cite{Ayon-Beato:2010vyw} are indeed recovered. As mentioned above, the present higher-curvature equations are much less restrictive than Einstein gravity. Notably, they admit solutions with a transverse section conformed by the product of any combination between spheres, and hyperbolic or flat planes; one such space for each NUT parameter. Of course, in this paper, our main concern has been the special case where they are all flat planes.

Conserved charges in quadratic gravity theories have been studied by using different techniques~\cite{Deser:2002rt,Deser:2002jk,Deser:2007vs,Baykal:2012rr,Giribet:2018hck,Giribet:2020aks,Miskovic:2022mqv,Miskovic:2023hzc}. However, the functional derivative of the Lagrangian with respect to the Riemann given by Eq.~\eqref{E-tensor} ---one of the key ingredients in all these methods--- vanishes when $R=4\Lambda$. This could indicate that, along the curve~\eqref{R2beta}, the theory might become critical whenever the Ricci scalar satisfies $R=4\Lambda$. Indeed, it is known that criticality avoids obtaining the energy formula from a linearized charges~\cite{Deser:2002jk,Camanho:2013pda,Fan:2016zfs,Petrov:2019roe,Arenas-Henriquez:2019rph} and going beyond the linear analysis is needed to obtain a definite answer. At this point, we are led to speculate that the conserved charges vanish, as in critical higher-curvature gravities. However, a definite answer can only be achieved by computing the conserved charges in full detail. This is certainly an interesting avenue to explore, but it lies beyond the scope of this paper. So, we postpone the analysis of the conserved charges of this solution to future work.

\section{Planar Kaluza-Klein multi-monopoles in AdS\label{sec:KKmonopoles}}

Before presenting the planar Kaluza-Klein multi-monopoles in AdS, let us construct a simple possible extension: the GPS Kaluza-Klein monopole~\cite{Gross:1983hb,Sorkin:1983ns} with cosmological constant. Although in~\cite{Mann:2005gk} the cosmological constant was introduced by fixing it in terms of the NUT charge, the construction presented here is devoid of such a restriction and admits a smooth flat limit when $\Lambda\to0$.

In the original Kaluza-Klein monopole~\cite{Gross:1983hb,Sorkin:1983ns}, the key idea is to identify the $U(1)$ fibration of the Euclidean Taub-NUT metric with the $U(1)$ structure that appears naturally in compactifications of $D+1$ dimensional spaces on $\mathbb{S}^1$. The Euclidean Taub-NUT metric, however, is four-dimensional, so it must be uplifted to five dimensions by adding a flat direction, i.e.,
\begin{align}\notag
    \diff{\bar{s}^2}_{\rm TN} &= \bar{g}_{AB}\diff{y^A}\diff{y^B} = \varepsilon\, \diff{z}^2 + f(r)(\diff{\psi} + 2n\cos\vartheta\diff{\varphi})^2 + \frac{\diff{r^2}}{f(r)} + (r^2-n^2)\diff{\Omega_{(2)}^2} \\
    &:= \varepsilon\,\diff{z^2} + g_{\mu\nu}\diff{x^\mu}\diff{x^\nu}\,, \label{TN5D}
\end{align}
where $\varepsilon=\pm1$ is introduced to analyze Euclidean and Lorentzian signatures on the same footing.\footnote{In the original GPS construction, a Lorentzian time was used, that is, $\varepsilon=-1$.} Here, capital Latin characters denote indices on the five-dimensional manifold with metric $\bar{g}_{AB}$, parametrized with local coordinates $\{y^A\}$, Greek characters denote indices on the four-dimensional Riemannian manifold with metric $g_{\mu\nu}$ and coordinates $\{x^\mu \}$, while $\diff{\Omega_{(2)}^2}$ denotes the line element of the round two-sphere.

The next step is to identify the uplifted Euclidean Taub-NUT metric in Eq.~\eqref{TN5D} with the $U(1)$ structure of the Kaluza-Klein ansatz in $D+1$ dimensions, that is,\footnote{Since there are several extensions of the Taub-NUT metric in higher dimensions (see, e.g. Ref.~\cite{Awad:2000gg}), this construction can be easily generalized to arbitrary dimensions.}
\begin{align}\label{KKansatz}
    \diff{\hat{s}^2_{\rm KK}} = \hat{g}_{AB}\diff{y}^A\diff{y^B} =  e^{2\alpha\phi}\diff{\tilde{s}^2} + e^{2\beta\phi}(\diff{\psi} + A)^2 \,,
\end{align}
where $\alpha^{-2} = 2(D-1)(D-2)$ and $\beta=-(D-2)\alpha$; although we will work in the case $D=4$ for simplicity. Additionally, $\phi=\phi(x)$ and $A=A_\mu\diff{x^\mu}$ are identified as the dilaton and Maxwell fields, respectively, depending only on the coordinates of the lower-dimensional metric $\diff{\tilde{s}^2}$. Direct comparison between Eqs.~\eqref{KKansatz} and~\eqref{TN5D} allows us to read the dilaton, Maxwell, and four-dimensional metric fields from the five-dimensional extension of the Taub-NUT metric, that is,
\begin{align}\label{mapTN-KK}
    \phi &=- \frac{\sqrt{3}}{2} \log f(r)\,, & A &= 2n\cos\vartheta\diff{\varphi}\,, & \diff{\tilde{s}^2} &= f(r)^{\frac{1}{2}}\left(\varepsilon\diff{z^2} + \frac{\diff{r^2}}{f(r)} + (r^2-n^2)\diff{\Omega_{(2)}^2} \right)\,, 
\end{align}
and $\chi=\lambda z$. Notice that the Maxwell field is exactly that of the Dirac monopole, as observed in~\cite{Gross:1983hb,Sorkin:1983ns}. However, in order to determine $f(r)$, we need to specify a higher-dimensional theory and solve its field equations, whose dimensional reduction will result in the lower-dimensional dynamics. 

The simplest case is to consider five-dimensional Einstein gravity. The vacuum field equations for the metric~\eqref{TN5D}, say $\bar{G}_{AB} + \Lambda \bar{g}_{AB}=0$, imply the Einstein equations for $g_{\mu\nu}$ along the $\mu\nu$ components, from which the metric function $f(r)$ can be obtained analytically. However, compatibility with the $zz$ components implies that the cosmological constant must be zero, obstructing spacetimes with $\Lambda\neq0$. However, as pointed out in Ref.~\cite{Cisterna:2017qrb}, a simple way to circumvent this restriction is to add a free scalar field $\chi$ that depends linearly on the coordinate that uplifts the four-dimensional metric to five dimensions, say $\chi=\lambda z$. This condition automatically solves the Klein-Gordon equation $\Box\chi=0$ on the background~\eqref{TN5D}. The Einstein equations, in turn, are now modified as $\bar{G}_{AB} + \Lambda\bar{g}_{AB}=\tfrac{1}{2}T_{AB}$, where the stress tensor of the scalar field is
\begin{align}
    T_{AB} = \nabla_A\chi\nabla_B\chi - \frac{1}{2}\bar{g}_{AB}(\nabla\chi)^2 \,.
\end{align}
The nontrivial components of the stress tensor are $T_{\mu\nu} =-\tfrac{1}{2}\varepsilon\lambda^2g_{\mu\nu}$ and $T_{zz}=\tfrac{1}{2}\lambda^2$. Thus, the $\mu\nu$ and $zz$ components of the Einstein equation can be written as~\cite{Cisterna:2017qrb}
\begin{subequations}\label{eomTN5Daxion}
    \begin{align}\label{eomTN5Dmunuaxion}
    R_{\mu\nu} - \frac{1}{2}g_{\mu\nu}R + \left(\Lambda + \frac{1}{4}\varepsilon\lambda^2 \right)g_{\mu\nu} &= 0\,, \\
    \varepsilon\left(-\frac{1}{2}R + \Lambda \right) &= \frac{1}{4}\lambda^2\,,
\end{align}
\end{subequations}
respectively. From here, it is clear that the absence of scalar fields ($\lambda=0$) implies that the cosmological constant must be zero. Remarkably, the vacuum constraint $\Lambda=0$ can be circumvented in $D+1$ dimensions if the integration constant $\lambda$ satisfies
\begin{align}\label{lambda2GPS}
    \lambda^2 = -\frac{4}{D-1}\varepsilon\Lambda\,.
\end{align}
Hence, the role of the free scalar field is to compensate for the curvature imbalance between Eqs.~\eqref{eomTN5Daxion}. Additionally, reality conditions on the scalar field impose that $\varepsilon\Lambda<0$. Therefore, the sign of the cosmological constant is determined by the signature of the metric~\eqref{TN5D}. Then, replacing Eq.~\eqref{lambda2GPS} into Eq.~\eqref{eomTN5Dmunuaxion}, one obtains $G_{\mu\nu} + \Lambda_{\rm eff}g_{\mu\nu}=0$, which represent the lower-dimensional Einstein equations with an effective cosmological constant given by
\begin{align}\label{Lambdaeff}
    \Lambda_{\rm eff}=\left(\frac{D-2}{D-1}\right)\Lambda\,.
\end{align}
Focusing on the AdS$_4$ case ($D=4$), that is, $\Lambda_{\rm eff}=-3/\ell_{\rm eff}^2$ and $\varepsilon=1$, the Euclidean Taub-NUT solution is given by
\begin{align}\label{fsolKKLambda}
    f(r) = \frac{r-n}{r+n} + \frac{\left(r-n\right)^{2}\left(r+3n\right)}{\ell^{2}_{\rm eff}\left(r+n\right)},
\end{align}
which is regular at $r=n$ as long as $\psi\sim\psi+8\pi n$. 

Since the previous construction utilizes the Einstein equations coupled to free scalar fields, the lower-dimensional theory can be obtained by performing the dimensional reduction of the Einstein-scalar theory using the Kaluza-Klein ansatz~\eqref{KKansatz}, giving
\begin{align}\label{EH5DKK}
    \sqrt{|\hat{g}|}\left(\hat{R}-2\Lambda-\frac{1}{2}(\nabla\chi)^2\right) = \sqrt{|\tilde{g}|}\left(\tilde{R} - 2\Lambda e^{\frac{\phi}{\sqrt{3}}} - \frac{1}{2}(\nabla\phi)^2  -\frac{1}{2}(\nabla\chi)^2 - \frac{1}{4}e^{-\sqrt{3}\,\phi} F^2  \right)\,,
\end{align}
where $F=\diff{A}$. Thus, the dimensional reduction of the Einstein-scalar action leads to the Einstein-dilaton-Maxwell theory with a Liouville potential and a free scalar field $\chi$.\footnote{The scalar $\chi$ can be coupled to the Pontryagin invariant of the Maxwell field, whose modified field equations are solved exactly without altering the original solution~\cite{Bardoux:2012aw}. However, this coupling cannot be generated from compactifications on $\mathbb{S}^1$ from odd to even dimensions, since the Pontryagin term exists only in $D=4m$ with $m\in\mathbb{Z}_{>0}$} The profiles~\eqref{mapTN-KK} solve the field equations of the right-hand side of Eq.~\eqref{EH5DKK} by construction, since the uplifted Euclidean Taub-NUT metric is a stationary point of the left-hand side. This solution has a smooth $\Lambda\to0$ (or $\ell_{\rm eff}\to\infty$) limit which, in turn, implies that $\lambda\to0$ by virtue of Eq.~\eqref{lambda2GPS}. This is a simple way to generalize the GPS monopole to include the cosmological constant.

To construct planar Kaluza-Klein multi-monopoles in AdS, we need to avoid the restriction imposed by the field equations, which require either that the cosmological constant vanish or that the NUT charges be equal. To do so, we take the solution in Sec.~\ref{sec:axions} as a seed metric and add a single flat direction, alongside a field with axionic profile depending linearly on the coordinate that spans the extra dimension, following the strategy of Ref.~\cite{Cisterna:2017qrb}. This allows us to uplift the even-dimensional solution to odd dimensions, keeping the cosmological constant nonzero. Concretely, we consider the ansatz in Eq.~\eqref{TN5D} with $g_{\mu\nu}$ being the analytic continuation of the line element~\eqref{ansatz} obtained by performing the Wick rotation $t\to-\ri\tau$ and $n_{(i)}\to -\ri n_{(i)}$. The Kaluza-Klein reduction is performed along the periodic coordinate that generates the $U(1)$ fibration over the Einstein-K\"ahler manifold in the seed metric. This procedure yields
\begin{align}\notag
    I = \int_{\mathcal{M}}\diff{^D}x\sqrt{|g|}\bigg[&R -2\Lambda e^{2\alpha\phi} - \frac{1}{2}(\nabla\phi)^2 - \frac{1}{4}e^{-2(D-1)\alpha\phi}\sum_{i=1}^kF^2_{(i)} \\ &\quad - \frac{1}{2}\sum_{i=1}^k|\nabla\varphi_{(i)}|^2 - \frac{1}{2}(\nabla\chi)^2\bigg]\,,\label{KKtheory}
\end{align}
where $A_{(i)}=2n_{(i)}B_{(i)}$ are the Abelian gauge fields that represent the planar Kaluza-Klein multi-monopoles with $n_{(i)}$ being the different magnetic charges, and $B_{(i)}$ is the K\"ahler potential one-form defined in Eq.~\eqref{Bi}. Additionally, $F^2_{(i)}=F_{\mu\nu}^{(i)}F^{\mu\nu}_{(i)}$, with $F_{(i)}=\diff{} A_{(i)}$ being the Abelian field strength. This is an Einstein-Maxwell-dilaton theory with a Liouville potential, minimally coupled to $k=(D-2)/2$ free complex scalar fields, $\varphi_{(i)}$, and one real scalar field $\chi$. The field equations are given by
\begin{subequations}\label{eomKK}
\begin{align}
    R_{\mu\nu} - \frac{1}{2}g_{\mu\nu}R + \Lambda e^{2\alpha\phi} &= \frac{1}{2}\bigg(T_{\mu\nu}^{(\varphi)} + T_{\mu\nu}^{(\phi)} + T_{\mu\nu}^{(\chi)} + e^{-2(D-1)\alpha\phi}T_{\mu\nu}^{(A)} \bigg) \,, \\
    \nabla_\mu\left(e^{-2(D-1)\alpha\phi}F^{\mu\nu}_{(i)} \right) &= 0\,, \\
    \Box\phi - 4\alpha\Lambda e^{2\alpha\phi} &= -\frac{1}{2}(D-1)\alpha e^{-2(D-1)\alpha\phi}\sum_{i=1}^k F_{(i)}^2 \,, \\
    \Box\varphi_{(i)} = 0\,, \quad \Box\bar{\varphi}_{(i)} &= 0\,, \quad \Box\chi =0\,, \label{eomaxionsKK}
\end{align}    
\end{subequations}
where the stress energy tensor for the complex scalar fields is given in Eq.~\eqref{Tmunuvarphi}, while for the dilaton and Abelian gauge fields, they are respectively given by\footnote{The stress tensor $T_{\mu\nu}^{(\chi)}$ can be obtained by replacing $\phi$ by $\chi$ in $T_{\mu\nu}^{(\phi)}$.}
\begin{subequations}
    \begin{align}
    T_{\mu\nu}^{(\phi)} = \nabla_\mu\phi\nabla_\nu\phi - \frac{1}{2}g_{\mu\nu}(\nabla\phi)^2\,, \quad \mbox{and} \quad 
    T_{\mu\nu}^{(A)} = \sum_{i=1}^{k}\left(F_{\mu\lambda}^{(i)}F^{(i)}_{\nu}{}^{\lambda} - \frac{1}{4}g_{\mu\nu}F_{(i)}^2 \right)\,.
\end{align}
\end{subequations}

In order to solve the field equations of the dimensionally reduced theory, the axionic contribution $\chi$ must be taken into account. The latter is necessary to ensure the existence of a nontrivial Liouville potential for the dilaton while maintaining a nonvanishing cosmological constant in the seed metric. This can be done by considering the following ansatz for the line element, free complex scalars, and axionic fields, that is,
\begin{align}\label{ansatzKK}
    \diff{s^2} &= f(r)^{\frac{1}{D-2}}\bigg(\varepsilon\,\diff{z^2} + \frac{\diff{r^2}}{f(r)} + \sum_{i=1}^{k}(r^2-n_{(i)}^2)\,\diff{\Sigma_{(i)}^2} \bigg)\,, \quad \varphi_{(i)} = \lambda_{(i)}\,z_{(i)}\,, \quad \chi = \lambda z\,,
\end{align}
respectively, where $\diff{\Sigma_{(i)}^2}$ is defined in Eq.~\eqref{dSigmai}, and $\lambda$ is a constant. The complex scalars and axionic field in Eq.~\eqref{ansatzKK} solve their associated field equations automatically. The remaining equations are solved by the metric function and the dilaton field
\begin{subequations}
    \begin{align}\label{fsolKK}
    f(r) &= -\frac{r}{\bar{P}_{(k)}(r)}\left(m+2\int^r\diff{\rho}\left[\Lambda_{\rm eff}  + \frac{1}{2}\sum_{i=1}^k\frac{\lambda_{(i)}^2}{\rho^2-n_{(i)}^2}\right]\frac{ \bar{P}^2_{(k)}(\rho)}{\rho\,\bar{P}'_{(k)}(\rho)}\right)\,, \\ \phi(r) &= -\frac{\ln f(r)}{2(D-2)\alpha}\,,
\end{align}
\end{subequations}
where $m$ is an integration constant, as long as the constraints~\eqref{lambda2GPS} and
\begin{align}\label{KKconstraint}
   \lambda_{(i)}^2 =\lambda_{(k)}^2 - \frac{2\Lambda_{\rm eff}}{k} \left(n_{(k)}^2-n_{(i)}^2\right)\,, \quad \forall\, i\;\; \mbox{with $k$ fixed},
\end{align}
are met. Notice that this constraint differs from that in Eq.~\eqref{lambdaconstraint} in a sign flip in front of the NUT charges; this difference arises after the Wick rotation $n_{(i)}\to -\ri n_{(i)}$ is performed. Similarly, $\bar{P}_{(k)}(r)$ denotes the Wick rotated function in Eq.~\eqref{Pdef}, that is,
\begin{align}
     \bar{P}_{(k)}(r) := \prod_{i=1}^k(r^2-n_{(i)}^2)\,, \quad \mbox{such that} \quad (\ln \bar{P}_{(k)})' = 2r\sum_{i=1}^k \frac{1}{r^2-n_{(i)}^2}\,.
\end{align}
In order to avoid a signature change in the metric~\eqref{ansatzKK}, the radial coordinate should be restricted to $r\geq r_h\geq n_{(i)}$, where $r_h$ is defined as the largest root of $f(r)$ in Eq.~\eqref{fsolKK}. Additionally, the coordinate $z$ has to be extended along the real line. Otherwise, one cannot eliminate conical singularities at $r=r_h$, and the axion $\chi$ would have been multi-valued. On the other hand, the behavior of the dilaton is similar to that in Ref.~\cite{Mann:2005gk}. However, the main difference is that the planar AdS Kaluza-Klein multi-instanton presented here has a smooth limit as $\Lambda\to0$ which, in turn, is translated into $\chi\to0$ according to Eq.~\eqref{lambda2GPS}, while keeping $n_{(i)}\neq 0$, with or without complex scalar fields $\varphi_{(i)}$. This is the main advantage of this construction when compared to that in Ref.~\cite{Mann:2005gk}.

\section{Conclusions\label{sec:conclusions}}

In this work, we have studied different scenarios for constructing planar multi-NUT configurations with a nonvanishing cosmological constant in higher dimensions. Previous work had shown that this was not possible in Einstein-AdS gravity due to strong constraints imposed by the field equations~\cite{Mann:2003zh,Mann:2005ra}. To do so, we consider two simple extensions that circumvent those restrictions. The first one is based on complex scalar fields with axionic profiles that depend linearly on the coordinates that span the flat transverse sections~\cite{Bardoux:2012aw}. The second one is inspired by quadratic-curvature corrections to the Einstein-Hilbert action, where higher-dimensional Lifshitz black holes have been found~\cite{Ayon-Beato:2010vyw}. These complementary directions represent the two main ways in which the vacuum Einstein field equations are modified: i) introducing matter fields and ii) including higher-curvature terms.

A natural application of our result is the planar Kaluza-Klein multi-monopoles with a nonvanishing cosmological constant~\cite{Mann:2005gk}. Following the GPS construction~\cite{Gross:1983hb,Sorkin:1983ns}, the even-dimensional solutions are uplifted by adding an extra flat direction, while keeping a nonzero cosmological constant via the mechanism of Ref.~\cite{Cisterna:2017qrb}. After performing the dimensional reduction along the periodic direction that generates the $U(1)$ fibration on the seed metric, we obtain a plane-symmetric Kaluza-Klein multi-monopole with $\Lambda\neq0$, whose vanishing cosmological constant limit keeps the magnetic charges turned on. What is more, by following this approach, we have generalized the GPS monopole, allowing for the inclusion of a cosmological constant. 

Several questions remain open and could lead to future research directions. Firstly, a detailed thermodynamic analysis of planar multi-NUT spaces is certainly worth exploring. The thermodynamics of NUT-type solutions introduces a long-studied difficulty, since the NUT charge and its conjugate variable in the first law of thermodynamics do not admit an independent derivation. One possibility to resolve this issue is to adopt the Bekenstein-Hawking area formula for the entropy, since the Misner string is absent and therefore does not modify it. This would allow one to derive the pair of thermodynamic quantities associated with the NUT charge and its conjugate potential, following the lines of Ref.~\cite{mann_thermo}. In the context of multi-NUT solutions, the first law of thermodynamics would be generalized to include multiple NUT charges and scalar charges associated with the translation symmetry. Only once the thermodynamic quantities have been identified can the thermodynamic ensemble be fixed and phase transitions explored. This would reveal possible phase transitions between these planar configurations and thermal AdS, which are otherwise forbidden in the absence of NUT charges. Secondly, The criticality of the quadratic theory where we have found multi-NUT spacetimes has not been established, and the conserved charges of those systems remain without analyzing. Moreover, it is known that the NUT charge in four dimensions can be interpreted as a magnetic mass~\cite{Hawking:1976jb,Hawking:1996ww,Hawking:1998ct,Lynden-Bell:1996dpw,Araneda:2016iiy,Corral:2024lva}. It would be interesting to study whether the multi-NUT spaces still admit the same type of interpretation in higher dimensions or if they induce some sort of angular momentum, similar to the Myers-Perry black hole. Thirdly, the role of multi-NUT parameters in fluid/gravity correspondence is certainly worth exploring, to see whether these configurations induce interesting effects in holographic fluid dynamics. It is worthwhile mentioning that after applying the large-gauge transformation of Ref.~\cite{Awad:2002cz} to the multi-NUT spacetimes presented here, the maximum number of rotations and NUT parameters would be turned on, hence allowing for any holographic fluid exploration involving plane symmetry. Such multi-NUT spacetimes also complement previous constructions with products of spheres as transverse sections~\cite{Chen:2006ea,Chen:2006xh}. Finally, much like Myers-Perry black holes, no electromagnetically charged multi-NUT spacetimes are known. In spite of the fact that charged Taub-NUT spacetimes are known in all dimensions~\cite{Awad:2005ff} and even in all Lovelock gravities~\cite{Corral:2025yvr}, which is not the case for Kerr-type systems. We leave these and other questions for future work. 

\section*{Acknowledgments}

We thank Eloy Ay\'on-Beato, Adolfo Cisterna, Felipe D\'iaz, Borja Diez, Rodrigo Olea, and Julio Oliva for helpful discussions. Additionally, we also thank the anonymous referee for their valuable comments and suggestions. The work of CC is partially supported by Agencia Nacional de Investigación y Desarrollo (ANID) through Fondecyt Regular grants No. 1240043, 1240048, 1252053, 1251523, and 1261016. DFA would like to acknowledge financial support from SECIHTI through a postdoctoral research grant. BH acknowledges the support of Direcci\'on de Postgrado, Universidad de Concepción, and FONDECYT Regular grant No 1240043.

\bibliographystyle{apsrev4-2}
\bibliography{References}

\end{document}